\documentclass[12pt,titlepage,a4paper]{article}
\usepackage{fancyheadings}
\usepackage{epsfig}
\usepackage{amssymb}
\usepackage{amsmath}
\usepackage{amsfonts}
\usepackage{color,graphicx}

\title {\sc {\bf A non-linear dynamical systems approach to source compression for constrained sources}}
\author{Nithin Nagaraj \\School of Natural and Engineering Sciences \\ National Institute of Advanced Studies\\ {\bf nithin@nias.iisc.ernet.in}  \\ \\  Prabhakar G Vaidya \\ School of Natural and Engineering Sciences \\ National Institute of Advanced Studies\\ {\bf pgvaidya@nias.iisc.ernet.in}\\ \\ Rajesh Sundaresan \\ ECE Dept., Indian Institute of Science\\ {\bf rajeshs@ece.iisc.ernet.in}}
\date {September 4, 2007} 
\begin{document}
\maketitle
\begin{abstract}
We have recently established a strong connection between the Tent
map (also known as Generalized Luroth Series or GLS which is a
chaotic, ergodic and lebesgue measure preserving non-linear
dynamical system) and Arithmetic coding which is a popular source
compression algorithm used in international compression standards
such as JPEG2000 and H.264.  This was for independent and
identically distributed binary sources. In this paper, we address
the problem of compression of ergodic Markov binary sources with
certain words forbidden from the message space. We shall show that
GLS can be modified suitably to achieve Shannon's entropy rate for
these sources.
\end{abstract}

\section{Introduction}
Emerging trends in data acquisition and imaging technologies have
resulted in a rapid increase in the volume of data. Hence source
coding (or data compression) continues to occupy an important area
of research in the design of communication and storage systems.
Shannon, the father of information theory, provided the definition
of Entropy as the ultimate limit of lossless data compression. Ever
since, there have been a number of compression algorithms which
tries to achieve this limit.

The source coding problem is stated as follows: Given an independent
and identically distributed (i.i.d) binary source $X$ emitting bits
of information in the absence of noise, how do we obtain the
shortest lossless representation of this information?  Source coding
is also known as entropy coding or data compression and is an
important part of most communication systems~\cite{Sayood96}.

Shannon in his 1948 masterpiece~\cite{Shannon48} defined the most
important concept of Information Theory, namely `Entropy'. Shannon's
Entropy of a source $H(X)$ is defined as the amount of information
content or the amount of uncertainty associated with the source, or
equivalently the least number of bits required to represent the
information content of a source without any loss. Shannon proposed a
method (Shannon-Fano coding~\cite{Salomon00}) that achieves this
limit as the block-length (number of symbols taken together) for
coding increases asymptotically to infinity.
Huffman~\cite{Huffman52} proposed what are called minimum-redundancy
codes with integer code-word lengths that achieve Shannon's Entropy
in the limit of the block-length tending to infinity. However, there
are problems associated with both Shannon-Fano coding and Huffman
coding (and other similar techniques). As the block-length
increases, the number of alphabets exponentially increase, thereby
increasing the memory needed for storing and handling. Also, the
complexity of the encoding algorithm increases since these methods
build code-words for all possible messages for a given length
instead of designing the codes for a particular message at hand.
Another disadvantage of all such methods is that they do not lend
themselves easily to an adaptive coding technique~\cite{Sayood96}.
The idea of adaptive coding is to update the probability model of
the source during the coding process. Unfortunately, for both
Huffman and Shannon-Fano coding, the updating of the probability
model would result in re-computation of the code-words for all the
symbols which is an expensive process.

Recently, we have proposed a new approach to address the source
coding problem using a dynamical systems
perspective~\cite{Nagaraj06_2}. We modeled the information bits of
the source $X$ as measurements of a non-linear dynamical system.
Since measurement is rarely accurate, we treat these measured bits
of information as a symbolic sequence~\cite{Nithin06_1} of the Tent
map~\cite{Alligood97} and their skewed cousins. Source coding is
seen as determination of the initial condition that has generated
the given symbolic sequence. Subsequently, we established that such
an approach leads us to a well known entropy coding technique
(Arithmetic Coding) which is optimal for compression. Furthermore,
this new approach enabled a robust framework for joint source coding
and encryption.

In this paper, we focus on constrained sources where certain words
are forbidden from its message space. Such sources are longer i.i.d
because they violate the independence assumption (most natural
sources are not independent, for e.g., english text is clearly not
independent). We consider only ergodic Markov sources in this paper.
We propose a non-linear dynamical systems approach to compress
messages from these constrained sources.

\section{Embedding an i.i.d source}
We shall first start with an i.i.d source $X$ which can be thought
of as emitting a sequence of random variables $\{X_1, X_2, \ldots,
X_N \}$. Each random variable takes values independently from the
common alphabet $\mathcal{A}= \{ a_1, a_2, \ldots, a_{|\mathcal{A}|}
\}$ with probabilities $\{p_1, p_2, \ldots, p_{|\mathcal{A}|} \}$
respectively. A message from this source is a particular sequence of
values $M=\{x_1, x_2, \ldots, x_N \}$ where $x_1$ is drawn from
$X_1$, $x_2$ from $X_2$ and so on. Since these are i.i.d, we can
think of them as being drawn from the common random variable $X$ (we
use the notation $X$ for both the source and the common random
variable). We always deal with finite sized alphabets
($|\mathcal{A}| < \infty$) and finite length messages ($N < \infty$)
since real-world messages are always finite in length.

\par Our aim is to embed this i.i.d source into a non-linear
discrete dynamical system. The reason for doing this will be clear
soon. To this end, we model the i.i.d source as a 0-order Markov
source (memoryless source). Each alphabet can be seen as a Markov
state. Since these are independent, the transition probability from
state $i$ to state $j$ is equal to the probability of being in state
$j$. In other words, $P_{ij} = P(X_{r+1}=j | X_r =i) =
P(X_{r+1}=j)$.

\par We wish to embed this 0-order Markov source into a non-linear
dynamical system $(\Omega,\Im , T, \mu)$ where $\Omega$ is the set
$[0,1)$, $\Im$ is the Borel $\sigma$-algebra on $[0,1)$, $T$ is the
measure preserving transformation (yet to be defined) and $\mu$ is
the invariant measure. Here we consider the probability measure (or
Lebesgue measure) as the invariant measure. We now need to define
$T$ which preserves the Lebesgue measure and which can {\it
simulate} the 0-order Markov source faithfully.

\subsection{GLS embeds the i.i.d source}
\begin{figure}[!h]
\centering
\includegraphics[scale=.5]{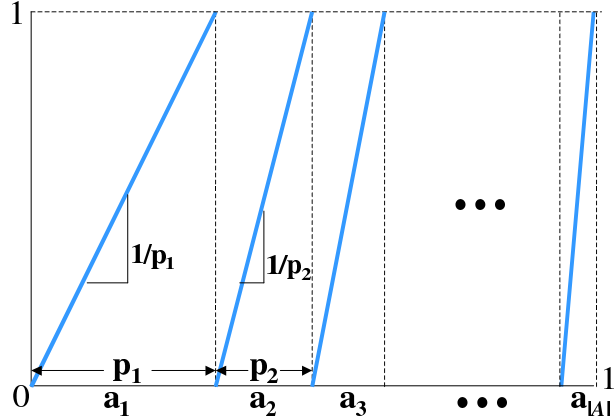}
\caption{GLS embeds an i.i.d source faithfully. The digits $\{ a_1,
a_2, \ldots, a_{|\mathcal{A}|} \}$  are the alphabets from
$\mathcal{A}$. The lengths of the intervals are precisely the
respective probabilities $\{p_1, p_2, \ldots, p_{|\mathcal{A}|}
\}$.} \label{fig:figGLS1}
\end{figure}

The non-linear discrete dynamical system known as Generalized
Lur\"{o}th series (GLS)~\cite{Dajani02} (Figure~\ref{fig:figGLS1})
embeds the 0-order Markov source. We list the important properties
of GLS which enable this embedding:
\begin{enumerate}
\item The number of partitions (disjoint intervals which cover the space, also known as cylinders)
of the GLS is equal to the size of the alphabet.
\item Each alphabet is used to `label' a partition.
\item The size of each partition is equal to the probability of the
corresponding alphabet.
\item The map is linear and surjective on each of the partitions.
\item GLS preserves the Lebesgue measure~\cite{Dajani02}.
\item Successive digits (or symbols) of the GLS are i.i.d.
\item Every unique sequence of digits (or symbols) maps to a
unique point $x$ in [0,1) under $T$. In other words, every point $x$
has an unique representation in terms of the alphabets of GLS. We
call $x$ as the initial condition corresponding to the symbolic
sequence (any sequence of digits composed from the alphabets
associated with the partitions).
\item GLS is chaotic (positive Lyapunov exponent and positive Topological entorpy).
\item The GLS transformation $T$ on [0,1) is isomorphic to the
Bernoulli shift~\cite{Dajani02}. Hence GLS is ergodic with Lebesgue
as the invariant measure.
\end{enumerate}

\subsection{Invariant distribution and Lyapunov exponent of GLS}
GLS preserves the Lebesgue measure. A probability density $\Pi(x)$
on [0,1) is invariant, if for each interval $[c,d] \subset [0,1)$,
we have:

\begin{equation*}
\int_c^d \Pi(x)dx = \int_{T^{-1}([c,d])} \Pi(x)dx.
\end{equation*}
where $T^{-1}([c,d]) = \{ x | c \leq T(x) \leq d \}$.

For the GLS, the above condition has constant probability density on
$[0,1)$ as the only solution. It then follows from Birkhoff's
ergodic theorem~\cite{Dajani02} that the asymptotic probability
distribution of the points of almost every trajectory is uniform. We
can hence calculate Lyapunov exponent as follows:

\begin{equation*}
\lambda = \int_0^1 log_2(|T'(x)|) \Pi(x)dx.~~~(a.e.)
\end{equation*}
Here, we measure $\lambda$ in bits/iteration.

\subsection{Shannon's entropy = Lyapunov exponent for GLS}
$\Pi(x)$ is uniform with value 1 on [0,1) and $T'(x) = constant$
since $T(x)$ is linear in each of the partitions, the above
expression simplifies to:

\begin{equation*}
\lambda = -\sum_{i=1,p_i\neq0}^{i=|\mathcal{A}|}
p_ilog_2(p_i).~~~(a.e.)
\end{equation*}
This is nothing but Shannon's entropy of the source $X$. Thus
Lyapunov exponent of the GLS that embeds the i.i.d source $X$ is
equal to the Shannon's entropy of the source. Lyapunov exponent can
be understood as the amount of information in bits revealed by the
dynamical system in every iteration(?). The number of partitions
together with the Lyapunov exponent completely characterizes GLS (up
to a permutation of the partitions and flip of the graph in each
parition - this changes the sign of the slope, but not its
magnitude).

\section{Coding the Initial Condition (GLS-coding)}
In the previous section, we have seen how we can embed a stochastic
i.i.d source $X$ in to a dynamical system (Generalized Lur\"{o}th
Series). The motivation for this is that modeling the stochastic
source by embedding in to a non-linear dynamical system is way to
achieve compression.



We know that Huffman coding is not Shannon optimal~\cite{Sayood96}.
This can be easily seen if the original source $X$ took only 2
values `a' and `b' with probabilities $\{p, 1-p\}$. For $p \neq
0.5$, Shannon's entropy of source $X$ is $<1$ bit whereas in Huffman
coding, we would allocate one bit to encode `a' and 'b'. That is the
best Huffman coding can do. Thus, by using Huffman coding, we would
be up to 1 bit away from Shannon's entropy per symbol. This can be
very expensive for skewed sources (where $p$ is close to 0 or 1)
which have a very low Shannon entropy. This means, we can do better
for such sources.

\par We have already said that every sequence of measurements (the
message) is a symbolic sequence on an appropriate GLS. It is well
known fact about dynamical systems that the symbolic sequence
contains as much information as the initial condition. Hence, we
could as well find out the initial condition for every symbolic
sequence and use that as our compressed stream. Hence, the task of
capturing the essential information of the source $X$ now translates
to determining the initial condition on the GLS (the source model)
and storing the initial condition in whatever base we wish
(typically the initial condition is binary encoded). Thus, the task
of source compression is now one of finding the initial condition.
We shall henceforth refer to this method as GLS-coding. How good is
GLS-coding when compared to Huffman coding?

\subsection{GLS-coding = Arithmetic Coding}
It turns out that the method just described is the popular
Arithmetic coding algorithm which is used in international
compression standards such as JPEG2000 and H.264. It is already
known that Arithmetic coding always achieves Shannon's optimality
without having to compute codewords for all possible messages and
this make it better than Huffman coding. We have thus re-discovered
Arithmetic coding using a dynamical systems approach to source
coding. For full details of the proof of equivalence with Arithmetic
coding, the reader is referred to \cite{Nagaraj06_2}.


\section{Constrained Source Coding (no longer i.i.d.)}
So far, we have dealt with an i.i.d source $X$. In the real world,
most sources are not independent even if they are identically
distributed. As an example, for a particle traveling in space, the
measurements of its position and velocity is clearly not independent
across successive time units. Thus, the assumption of independence
needs to be relaxed.

In communications, independence assumption is not generally true.
For example, assume that the source $X$ is an excerpt from an
English text. The probability that the letter `$u$' appears after
`$q$' is very high. Thus, given that a particular symbol has
occurred, there is probably a very small set of alphabets that can
occur with a very high probability.

In this paper, we consider another kind of dependence known as a
constrained source. A constrained source is one where certain
`words' are forbidden from the message space. As an example, if the
source is an English text, certain words are forbidden (words which
are profane or even words which are just gibberish, like for e.g.,
`QWZTY'). For the rest of the paper, we shall consider a binary
constrained source $X$. The constraint is given in terms of a list
of forbidden words (for e.g., the word `101' and `011' may be known
never to occur in any of the messages emitted by the source, then
both these words are defined as forbidden words). This is clearly
not an i.i.d source any more. Because given that a `10' as occurred,
the next symbol can only be `0'. Thus the present symbol depends on
what occurred in the last two instances. We are interested in
compressing sequences of such a source in the most optimal fashion.

\begin{figure}[!h]
\centering
\includegraphics[scale=.6]{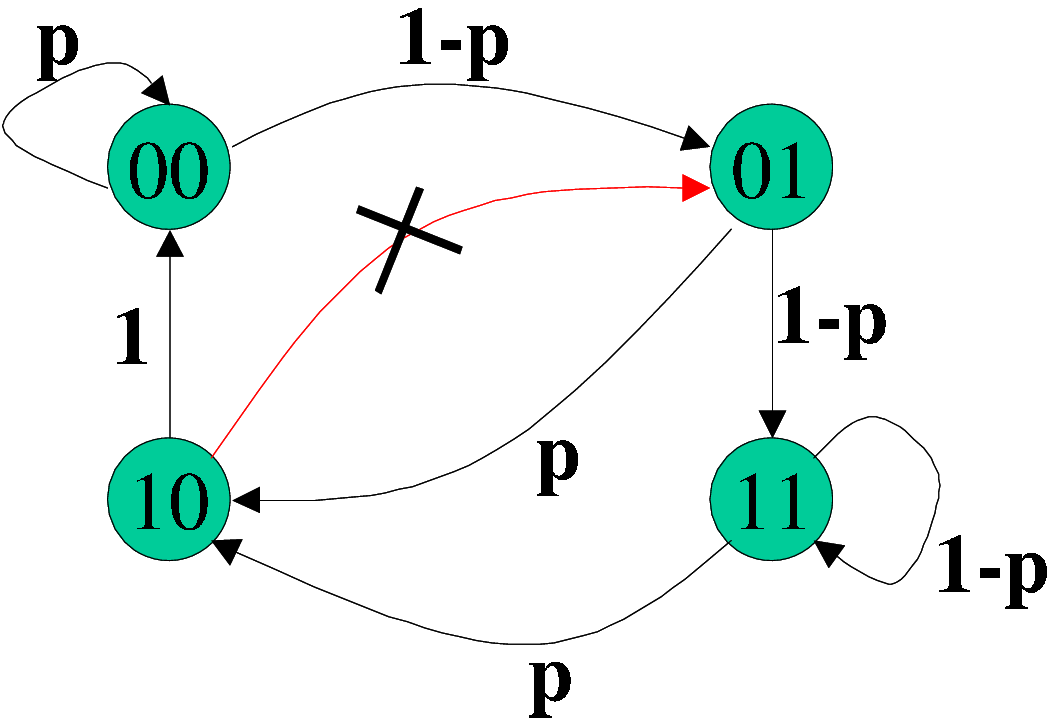}
\hspace{0.1in}
\includegraphics[scale=.6]{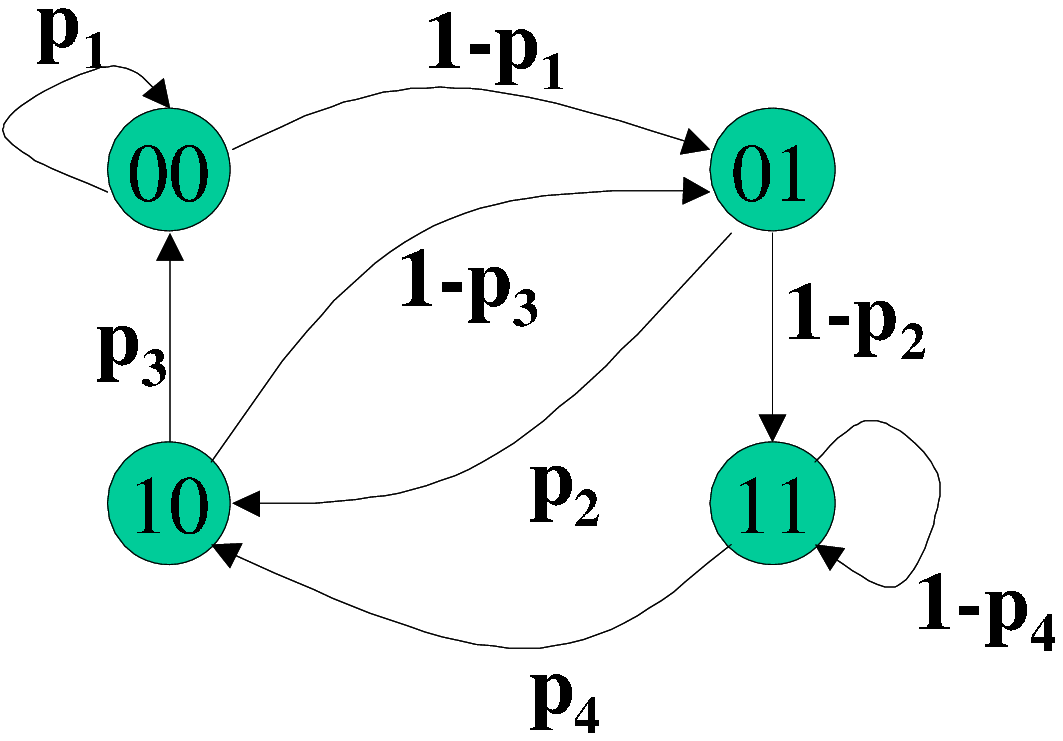}
\caption{(a) Left: Constrained ergodic Markov source (forbidden word
= `101'). (b) Right: A general 4-state ergodic Markov source. Note:
The case $p_1=p_2=p_3=p_4=p$ yields an i.i.d source and the case
$p_1=p_2=p_4=p, p_3=1$ yields the source with forbidden word `101'.
Thus the general 4-state ergodic Markov source captures both cases.}
\label{fig:figforbidden12}
\end{figure}

\subsection{Constrained Ergodic Markov Sources}
We shall model constrained sources as a Markov source. In this
paper, we shall consider only those Markov sources that are ergodic.
A Markov source is ergodic if either the Markov chain itself is
ergodic~\cite{Markov1} or equivalently, the dynamical system in
which the Markov chain is embedded is ergodic. We know by the theory
of Markov chains~\cite{Markov1} that all finite discrete time Markov
chains that are ergodic (i.e. they are irreducible and aperiodic)
have an unique stationary distribution, also known as invariant
distribution. This means that given any arbitrary initial
distribution on the Markov states, it eventually settles to an
unique probability distribution. Once we have an unique stationary
distribution with finite states, Shannon's entropy rate can be
calculated. In general, it is hard to determine the invariant
measure of the underlying dynamical system in which the Markov chain
is embedded, but it is relatively easy to determine whether the
Markov chain is ergodic (all we need to do is test for
irreducibility and aperiodicity).

We already saw in Section 2 how an i.i.d source $X$ can be seen as a
0-order Markov chain with finite number of states. We could do
similarly for ergodic Markov sources. As an example, a binary
ergodic Markov source with `101' as a forbidden word is shown in
Figure~\ref{fig:figforbidden12}.

\subsection{Computing Shannon's entropy rate}
In Figure~\ref{fig:figforbidden12}, a general four state Markov
source is also shown. The condition $p_1=p_2=p_3=p_4=p$ implies that
the source is i.i.d. The condition $p_1=p_2=p_4=p$ and $p_3=1$
corresponds to the ergodic Markov source with the forbidden word
`101'. We shall deal with the general four state Markov source
(assuming that it is ergodic).

\par The transition probability matrix for the general four state
ergodic Markov source is given by:
\begin{equation*}
P = \begin{bmatrix}
      p_1 & 1-p_1 & 0 & 0 \\
      0 & 0 & p_2 & 1-p_2 \\
      p_3 & 1-p_3 & 0 & 0 \\
      0 & 0 & p_4 & 1-p_4 \\
    \end{bmatrix}
.
\end{equation*}
\par The unique stationary probability distribution $\Pi$ (a row vector of dimension $1\times|\mathcal{A}|$) can be
determined by solving the equation:
\begin{equation}
\Pi P = P.
\end{equation}

\par Since $P$ is always a stochastic matrix (every row adds to 1) of dimension $|\mathcal{A}|\times|\mathcal{A}|$,
there exists a unique eigenvector $\Pi$ for $P$ corresponding to the
eigenvalue 1. We normalize $\Pi$ as follows:
\begin{equation}
\hat{\Pi} = \frac{\Pi}{\| \Pi \|}.
\end{equation}

\par Once we have $\hat{\Pi} = \{ \Phi_1, \Phi_2, \ldots, \Phi_{|\mathcal{A}|}
\}$ we can compute Shannon's entropy of the source $X$ as follows:

\begin{equation}
H_X = -\frac{1}{log_2(Num)}\sum_{i=1, \Phi_i \neq
0}^{i=|\mathcal{A}|} \Phi_i log_2(\Phi_i).
\end{equation}

The units of $H_X$ is bits/symbol and $Num$ represents the number of
Markov states. For the general 4-state ergodic Markov source, this
will be 1/2 since each states accounts for 2 bits of the message.

\subsection{Modified GLS-coding}
In this section, we shall embed the general 4-state ergodic Markov
source into a non-linear dynamical system. We shall determine the
initial condition for any given symbolic sequence (message) on the
resulting dynamical system and use the initial condition to compress
the message (similar to GLS-coding). We shall show that such a
method achieves Shannon's entropy rate $H_X$.

\begin{figure}[!h]
\centering
\includegraphics[scale=.6]{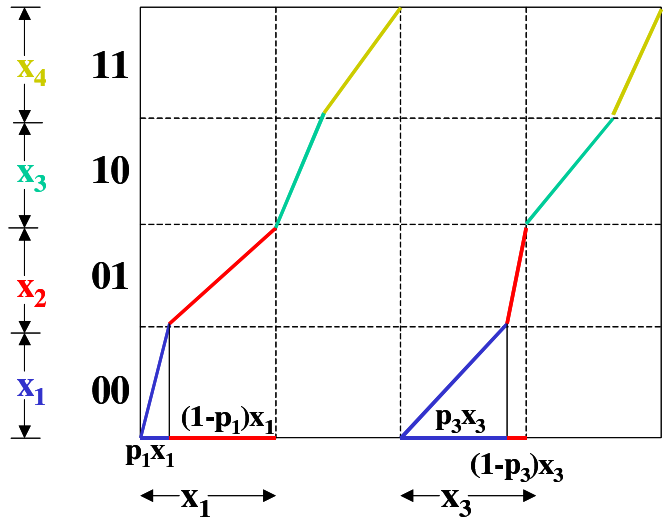}
\caption{Embedding a constrained ergodic Markov source in a
dynamical system with the map $T$:~[0,1) $\mapsto$ [0,1). }
\label{fig:figmodgls1}
\end{figure}

\subsection{Embedding}
We shall embed the general 4-state ergodic Markov source into a
non-linear dynamical system similar to GLS for the i.i.d source. To
each Markov state, we associate a Markov partition of the dynamical
system (refer to Figure~\ref{fig:figmodgls1}). We want the messages
of this source to be symbolic sequences of the dynamical system.
Draw straight lines with slopes yet to be determined connecting
those states that communicate. For example, 00 communicates only
with 01 and 10. Let the lengths of the Markov paritions be $x_1$,
$x_2$, $x_3$ and $x_4$. We then write the ``measure-preserving''
constraints as follows:

\begin{eqnarray*}
p_1x_1 + p_3x_3 &=& x_1 ~~~ (for~~00)\\
(1-p_1)x_1 + (1-p_3)x_3  &=& x_2 ~~~ (for~~01)\\
p_2x_2 + p_4x_4 &=& x_3 ~~~ (for~~10)\\
(1-p_2)x_2 + (1-p_4)x_4  &=& x_4 ~~~ (for~~11) \\
\end{eqnarray*}

These constraints automatically satisfy $x_1 + x_2 + x_3 + x_4 =1$.
Solving the first two of the above equation yields $x_3 = x_2$. The
above linear set of equations can be solved for the given set of $\{
p_1, p_2, p_3, p_4 \}$ (if the Markov chain is ergodic, a unique
solution always exists). The solution gives the length of the Markov
partitions $\{ x_1, x_2, x_3, x_4 \}$ which is unique. The slopes of
the line segments are determined from the probabilities. We thus
have a dynamical system (modified GLS) and we shall show that this
embeds the ergodic Markov source.

\subsection{Interpretation of the ``measure-preserving'' constraints}
How do we understand the ``measure-preserving'' constraints? As an
example, consider the Markov partition of length $x_1$ corresponding
to the state 00. Since 00 communicates only with itself and 01, we
have the range of the function limited to intervals 00 and 01. The
slope of the line segment (in blue) which maps fraction of $x_1$ to
00 is $\frac{1}{p_1}$. This is because whenever we are in state 00,
the probability that we end up in the same state on receiving the
next symbol is $p_1$. Thus $p_1x_1$ is the fraction of initial
conditions which end up in the same state 00. The remaining
$(1-p_1)x_1$ fraction of initial conditions end up in 01. Similarly,
we do this for all the states. The ``measure-preserving'' constraint
for say the state 00 is indicating what fraction of initial
conditions end up in state 00 in one iteration. This is formed
precisely as the sum of the fraction $p_1x_1$ which come from 00 and
$p_3x_3$ that comes from 10 (because these are the only two states
that communicate with 00). This is how we get all the constraint
equations.

\subsection{``Measure-preserving'' constraints are equivalent to $\Pi P = P$.}
The linear set of ``measure-preserving'' constraints can be
written in matrix form as follows:\\
\begin{equation*}
\begin{bmatrix}
   x_1 & x_2 & x_3 & x_4 \\
 \end{bmatrix}
 \begin{bmatrix}
  p_1 & 1-p_1 & 0 & 0 \\
  0 & 0 & p_2 & 1-p_2 \\
  p_3 & 1-p_3 & 0 & 0 \\
  0 & 0 & p_4 & 1-p_4 \\
\end{bmatrix} = \begin{bmatrix}
   x_1 & x_2 & x_3 & x_4 \\
 \end{bmatrix}.
\end{equation*}

\par Notice that the above equation is the same as $\Pi P = P$ where $P$ is
the transition probability matrix of the Markov source. Thus $\{
x_1, x_2, x_3, x_4 \}$ is nothing but the unique stationary
probability distribution obtained from the equation $\Pi P = P$.

\par We have thus embedded the ergodic Markov source in to a non-linear
dynamical system.

\begin{figure}[!h]
\centering
\includegraphics[scale=.6]{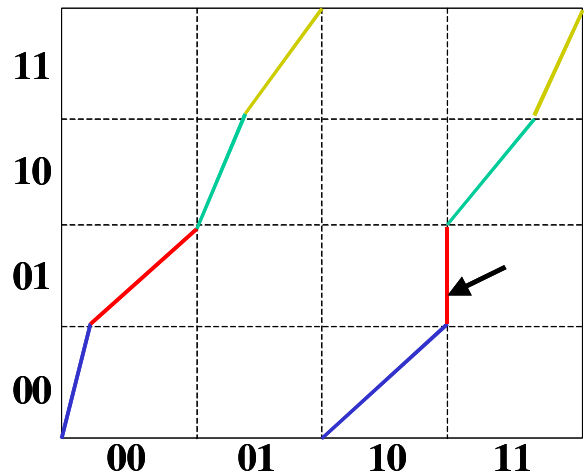}
\caption{Embedding for the source with forbidden word `101'. Notice
how one of the slope becomes infinite since `10' is forbidden to
communicate with `01'. } \label{fig:figmodgls2}
\end{figure}

\subsection{Modified GLS-coding achieves Shannon's entropy rate}
In order to prove that encoding the initial condition on the
modified GLS will achieve Shannon's entropy rate, we compute the
Lyapunov exponent of the modified GLS and show that this is the same
as Shannon's entropy rate. In fact, the non-linear dynamical system
is a faithful modelling of the general 4-state ergodic Markov
source.

\par It is important to observe that the modified GLS shown in
Figure~\ref{fig:figmodgls1} does not preserve the Lebesgue measure.
However, if we restrict all the intervals on the y-axis to strictly
lie in one of the four intervals labelled 00, 01, 10 and 11 only,
then we see that the Lebesgue measure (=probability measure) is
actually preserved (sum of the measures of all the inverse images of
a particular interval on the y-axis is equal to the measure of that
interval we started with). In fact, this is what we ensured by the
``measure preserving'' constraints in the first place. We could
hence compute the Lyapunov exponent as if the map preserved the
Lebesgue measure everywhere.

\par Computation of Lyapunov exponents is then straightforward and
yields the following expression (the base of the logarithm in all
our Lyapunov exponent computation is always 2, though the standard
practice is to use $e$.)

\begin{equation*}
\lambda = -\sum_{i=1}^{4} p_i log_2(p_i) = 2H_X.
\end{equation*}

This is nothing but twice the Shannon's entropy rate $H_X$ of the
source. The factor 2 is because in every iteration, the symbolic
sequence emitted by the source consists of two symbols. Thus, we
have faithfully modelled the general 4-state ergodic Markov source
as a non-linear dynamical system. Hence modified-GLS coding would
achieve the Shannon's entropy rate and is optimal for compression.




\section{Conclusions and Future Research Directions}
In this paper, we have shown how one can embed an i.i.d source in to
a non-linear dynamical system, namely Generalized Lur\"{o}th Series
or GLS. We then considered Markov sources which are not independent
anymore. Constrained sources were defined as ergodic Markov sources
with certain forbidden words. We showed how to compute the Shannon's
entropy rate and also modified GLS to compresses messages from these
sources. The modified-GLS is a faithful embedding of constrained
sources and hence achieves Shannon's entropy rate.

It is possible to generalize our method for a list of arbitrary
forbidden words. Implementation issues were not discussed in this
paper. It may be worthwhile to investigate joint compression and
encryption for constrained sources.

\section*{Acknowledgements}
Nithin Nagaraj would like to express his sincere gratitude to the
Department of Science and Technology (DST) for funding the Ph.D.
fellowship program at National Institute of Advanced Studies (NIAS).
We gratefully acknowledge DST, Govt. of India and Council of
Scientific and Industrial Research (CSIR), Govt. of India for
providing with travel grant to present this work at the
``International Conference on Non-linear Dynamics and Chaos:
Advances and Perspectives'', held at University of Aberdeen,
Scotland, September 17-21, 2007.

\end{document}